\documentclass[12pt]{article}
\def\bSig\mathbf{\Sigma}

\usepackage{amsmath}
\usepackage{graphicx,psfrag,epsf}
\usepackage{enumerate}
\usepackage{mathtools}
\usepackage{natbib}
\usepackage{url}
\usepackage{floatrow}
\usepackage{float}
\usepackage{bm}
\usepackage{ulem}
\usepackage[title]{appendix}

\newcommand{\blind}{1}

\begin{document}

\def\spacingset#1{\renewcommand{\baselinestretch}%
{#1}\small\normalsize} \spacingset{1}


\if1\blind
{
  \title{\bf Multivariate Hierarchical Frameworks for Modelling Delayed Reporting in Count Data}
  \author{Oliver Stoner\thanks{The authors gratefully acknowledge the Natural Environment Research Council for funding this work through a GW4+ Doctoral Training Partnership studentship [NE/L002434/1].}\hspace{.2cm}\\
    Department of Mathematics, University of Exeter\\
    and \\
    Theo Economou \\
    Department of Mathematics, University of Exeter}
  \maketitle
} \fi

\if0\blind
{
  \bigskip
  \bigskip
  \bigskip
  \begin{center}
    {\LARGE\bf Multivariate Hierarchical Frameworks for Modelling Delayed Reporting in Count Data}
\end{center}
  \medskip
} \fi

\bigskip
\begin{abstract}
In many fields and applications count data can be subject to delayed reporting. This is where the total count, such as the number of disease cases contracted in a given week, may not be immediately available, instead arriving in parts over time. For short term decision making, the statistical challenge lies in predicting the total count based on any observed partial counts, along with a robust quantification of uncertainty.

In this article we discuss previous approaches to modelling delayed reporting and present a multivariate hierarchical framework where the count generating process and delay mechanism are modelled simultaneously. Unlike other approaches, the framework can also be easily adapted to allow for the presence of under-reporting in the final observed count. To compare our approach with existing frameworks, one of which we  extend to potentially improve predictive performance, we present a case study of reported dengue fever cases in Rio de Janeiro. Based on both within-sample and out-of-sample posterior predictive model checking and arguments of interpretability, adaptability, and computational efficiency, we discuss the advantages and disadvantages of each modelling framework.
\end{abstract}

\noindent%
\vfill

\newpage
\spacingset{1.5} 

\section{Introduction}\label{sec:intro}
In many fields and applications where count data are collected, a situation can arise where the available reported count is believed to be less than or equal to the true count. Delayed reporting in particular is where the total observable count, which may still be less than the true count, is only available after a certain amount of time. In some situations information will trickle in over time so that the current total count gets ever closer to the true count, before eventually reaching the final total observable count.
\begin{figure}[h!]
\floatbox[{\capbeside\thisfloatsetup{capbesideposition={right,center},capbesidewidth=0.4\linewidth}}]{figure}[1.15\FBwidth]
{\caption{Bar plot of Rio de Janeiro dengue cases in the weeks leading up to time $t=114$. The grey bars represent the total (as yet unobserved) number of reported cases, while the coloured bars show the number of cases reported in each week after the cases occurred.} \label{fig:dr:bar_plot}}
{\includegraphics[width=\linewidth]{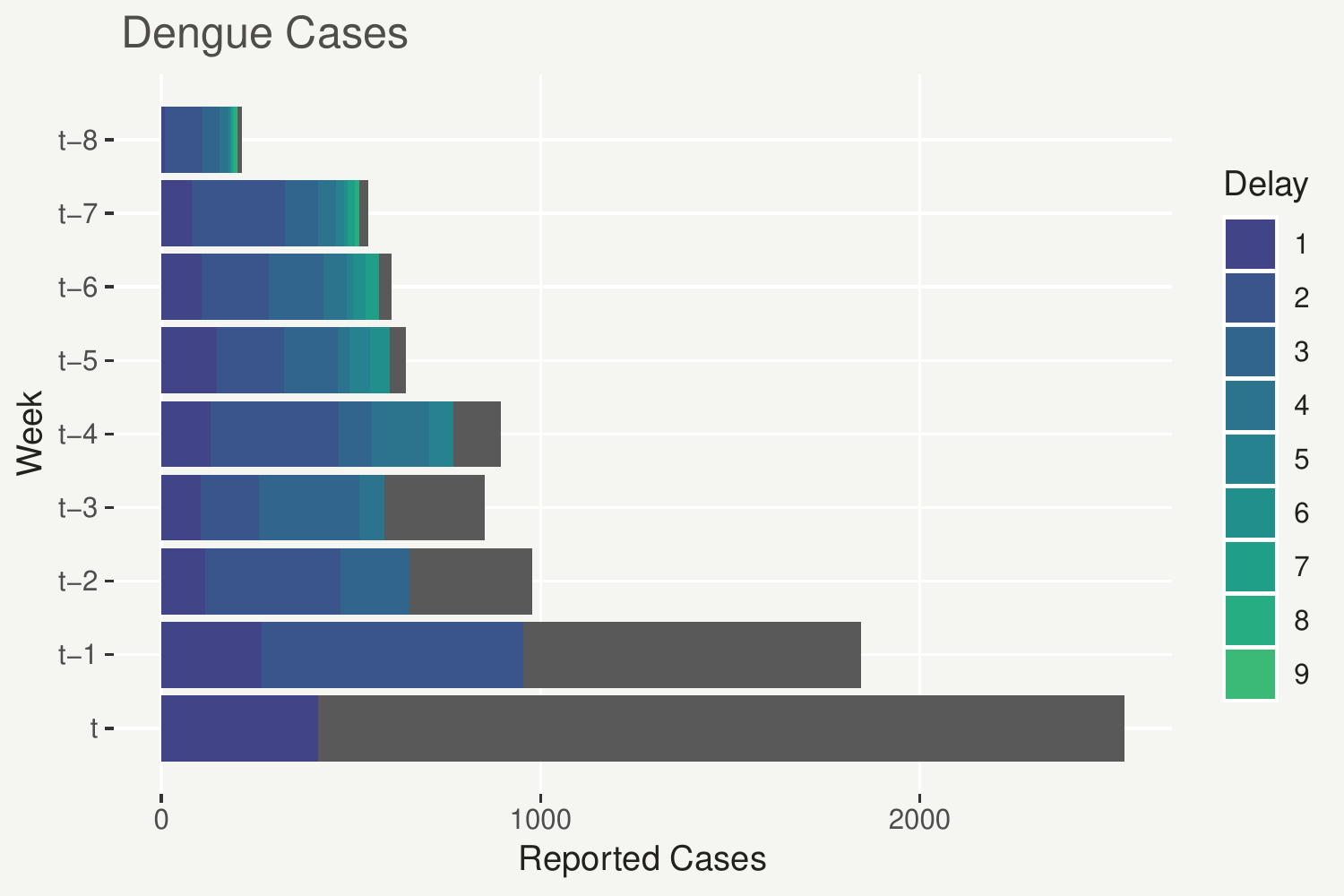}}
\end{figure}

An example of this situation is the occurrence of dengue fever, a viral infection spread by mosquitoes, in Rio de Janeiro. Imagining we are at the end of week $t$, due to delayed reporting we have only observed a portion of the total observable number of cases which were contracted this week. A week from now, at time $t+1$, a further portion will become available and so on, such that after a number of weeks the total number of observed cases we have observed from week $t$ eventually reaches a final total. Figure \ref{fig:dr:bar_plot} shows an instance of the data, where we are at the end of week $t=114$. The grey portions of each bar represent the yet unknown cases as of week $t$, for instance we can see that for dengue cases that occurred in week $t-1$ we only have two weeks worth of information because we only have information that arrived in weeks $t-1$ and $t$, while for cases occurring in week $t-2$ we have three weeks worth of information and so on.

Reporting delay becomes a problem when decisions need to be made based on the total count before it has been observed in its entirety. We can see in Figure \ref{fig:dr:bar_plot}, for example, that in the surveillance of dengue fever it can take months before the total observable number of cases contracted in a given week is known. This impedes the response time to severe outbreaks and puts lives at risk. It is therefore necessary to make predictions about the current state of the disease based on the partial number of cases observed (now-casting). This allows warnings to be issued and preparations made for predicted epidemics before they have been completely detected by the data. This motivates a statistical treatment of delayed reporting, with the goal of being able to predict the total count based on corresponding counts already observed. Further goals include predicting total counts which have not occurred yet (forecasting) and learning about the structure of the delay mechanism, so that improvements in reporting can be considered.

In this article we explore previous statistical approaches to modelling delayed reporting in count data, and discus their strengths and weaknesses. We then propose a general framework for modelling count data with discrete-time delays, which is sufficiently flexible to be used for a range of data, including those with complex spatio-temporal structures, and can be easily adapted to account for the presence of under-reporting in the final observed count. We present a case study based on counts of dengue fever cases in Rio de Janeiro, and assess the proposed framework by posterior predictive checking (now-casting and forecasting performance), relative to existing approaches.

The article is structured as follows: Section \ref{sec:background} presents an overview of existing approaches to modelling count data suffering from discrete-time delayed reporting, in addition to a substantial extension to one of the existing approaches. In Section \ref{sec:framework} we propose a general framework for modelling delayed reporting. In Section \ref{sec:comparison} we present a case study of dengue fever data from Rio de Janeiro, with which we assess the efficacy of our framework compared to existing approaches. In Section \ref{sec:under-reporting}, we discuss the potential issue of under-reporting in the final observed count and how the general framework from Section \ref{sec:framework} can be adapted to account for it. Finally, Section \ref{sec:discussion} concludes with a discussion of interpretability, adaptability and ease of implementation.

\section{Background}\label{sec:background}

We begin by introducing some notation. Let $y_{t,s}$ be the total observable count occurring at temporal unit $t \in T$ and spatial unit $s \in S$. We refer to $y_{t,s}$ as the total observable count because, in some cases, the final count we observe may still be an under-representation of the true count, an issue we return to in Section \ref{sec:under-reporting}. Suppose that after some (temporal) delay unit (e.g. one week) a portion of $y_{t,s}$ has been reported. We denote this first portion $z_{t,s,1}$. At the next delay unit we observe an additional portion of $y_{t,s}$, denoted as $z_{t,s,2}$. This continues so that at each delay unit $d \in \{1,...,D\}$ we observe a count $z_{t,s,d}$ so that the sum of the observed $z_{t,s,d}$ gets closer to the total count $y_{t,s}$. 
\subsection{Multinomial mixture approach}
A sensible approach for modelling delayed reporting involves the idea of jointly modelling $z_{t,s,d}|y_{t,s}$ at the same time as the totals $y_{t,s}$. \cite{hohle2014bayesian} propose modelling the delayed counts $\bm{z}_{t,s}|y_{t,s}$ as arising from a conditional Multinomial$(\bm{p}_{t,s},y_{t,s})$ distribution. Here $p_{t,s,d}$ is the expected proportion of $y_{t,s}$ which will be reported at delay $d$ and is modelled as arising from Generalized-Dirichlet($\bm{\alpha},\bm{\beta}$) (GD) distribution \citep{generalizeddirichlet} where $\bm{\alpha}$ and $\bm{\beta}$ are constant in time. The total observable count is also modelled explicitly as a latent Poisson variable in the Multinomial model:
\begin{eqnarray}
y_{t,s} \mid \lambda_{t,s} &\sim & \mbox{Poisson}(\lambda_{t,s}) \\
\bm{z}_{t,s} \mid \bm{p}_{t,s},y_{t,s} &\sim & \mbox{Multinomial}(\bm{p}_{t,s},y_{t,s}) \\
\bm{p}_{t,s} \mid \bm{\alpha},\bm{\beta} &\sim& \mbox{Generalized-Dirichlet}(\bm{\alpha},\bm{\beta})
\end{eqnarray}
\cite{foodborne} also apply this approach to the monitoring of foodborne diseases, while a similar approach (without the General-Dirichlet layer) can be found in \cite{salmon1}.

However, the assumption that the Generalized-Dirichlet distribution is time-invariant can be viewed as a restriction in capturing any delay mechanism which varies systematically over time, potentially inhibiting nowcasting and forecasting precision. \cite{hohle2014bayesian} seek to address this by presenting a second approach in which the Generalized-Dirichlet model is replaced with a more conventional logistic regression on the Multinomial probabilities:
\begin{eqnarray}
\log\left(\frac{\nu_{t,s,d}}{1-\nu_{t,s,d}}\right) &=& g(t,s,d) \\
p_{t,s,d} &=& \nu_{t,s,d}\left(1-\sum_{i=1}^{d-1}p_{t,s,i}\right)
\end{eqnarray}
where $g(t,s,d)$ is a linear combination of covariate effects. However, whilst this does allow the model to better capture heterogeneity in the delay mechanism over time, it is in part more restrictive. This is because in some applications the Multinomial delay model may be over-dispersed. We will discuss this issue in more detail in Section \ref{sec:framework}, where we propose a general framework which retains both the flexibility to capture spatio-temporal heterogeneity as well as the ability to appropriately separate variability in the delay mechanism from the model of the total count.

\subsection{Conditional independence approach}\label{sec:glm}
An alternative approach, often used in the field of actuarial statistics for projecting ultimate losses from delayed insurance claims, is the Chain-Ladder method \citep{mack_1993}. The method is popular because it is easy to understand and is based entirely on empirical calculations. \cite{renshaw_1998} showed that the Chain-Ladder method can be presented as a Generalized Linear Model \citep{GLM} of the following form:
\begin{eqnarray}
z_{t,d} & \sim & \mbox{Poisson}(\mu_{t,d}) \\
\log ( \mu_{t,d} ) &=& \iota + \alpha_t + \beta_d
\end{eqnarray}
This has been extended in various ways, for example to include potential covariates (see for instance \cite{england_verrall_2002} and \cite{BARBOSA2002}). These approaches however, are restrictive in the sense that they assume the delay structure is homogeneous in time. In reality, the way in which reporting is delayed, for example the proportion of cases reported in the first week, changes over time. The baseline Chain-Ladder model has therefore been extended to accommodate such non-homogeneities as well as spatial variability.

A highly flexible approach that in some sense generalises the Chain-Ladder, is the conditional independence approach where the partial counts $z_{t,s,d}$ ($d \in \{1,...,D\}$) are modelled as independent quantities, conditional on any spatio-temporal or delay structures in their expected value. We refer to this as the Generalized Linear Model (GLM) approach, as it is effectively (conditional on dispersion parameters) a Negative-Binomial GLM \citep{GLM} for the partial counts $z_{t,s,d}$:
\begin{eqnarray}
z_{t,s,d} \mid \mu_{t,s,d}, \theta &\sim & \mbox{Neg-Bin}(\mu_{t,s,d},\theta) \label{eq:cia1}\\
\log(\mu_{t,s,d}) &=& f(t,s)+g(t,s,d) \label{eq:cia2}
\end{eqnarray}

Here $f(t,s)$ and $g(t,s,d)$ can be linear combinations of covariate effects or random effects, including complex temporal, spatial and spatio-temporal structures. The former is intended to capture variation in the total counts $y_{t,s}$ while the latter is intended to capture variation in the delay mechanism. Aside from the flexibility of incorporating complex structures in the model for $\mu_{t,s,d}$, the key advantage of this approach is that it can be very easily implemented in a variety of frequentist frameworks (such as Generalized Additive Models, \cite{GAM}), as well as Bayesian ones (such as Integrated Nested Laplacian Approximations (INLA) \citep{r-inla} and Markov Chain Monte Carlo (MCMC)). For example, \cite{theo_dengue} presents the application of this framework to dengue fever in Rio de Janeiro and to spatio-temporal Severe Acute Respiratory Infection (SARI) data in the state of Paran\'a (Brazil). Both were implemented in the Bayesian framework using INLA and in this case the framework was demonstrated to be a powerful tool for now-casting. 

However, as $y_{t,s}$ is not modelled directly, inference is based on $y_{t,s}=\sum_{d=1}^{D} z_{t,s,d}$. Firstly, this means that uncertainty associated with the delay component of the GLM is potentially transferred through the summation of the $z_{t,s,d}$ into the uncertainty of the $y_{t,s}$. A consequence of this uncertainty propagation is that models such as \eqref{eq:cia1}-\eqref{eq:cia2} may result in forecasting uncertainty (for example as quantified by 95\% prediction intervals) that is prohibitively large, particularly when forecasting into the future where no $z_{t,s,d}$ are available. 

Furthermore, to obtain reliable inference for $y_{t,s}$, we would expect the model to capture $\mbox{Var}(y_{t,s})$ well. As $y_{t,s}$ is not modelled directly this is given by:
\begin{equation}\label{varY}
\mbox{Var}[y_{t,s}]=\mbox{Var}\left[\sum_{d=1}^{D} z_{t,s,d}\right]=
\sum_{i=1}^{D}\sum_{j=1}^{D} \mbox{Cov}[z_{t,s,i},z_{t,s,j}]
\end{equation}
This means that capturing the variance of $y_{t,s}$ well relies on modelling the covariances of the $z_{t,s,d}$ well. The issue with this is that the covariances of the $z_{t,s,d}$ are restricted by the assumption that $z_{t,s,d}$ are independent, conditional on $\mu_{t,s,d}$. In many cases this may not be a valid assumption and consequently any inference based on $y_{t,s}$ is fundamentally flawed. To potentially address this, in the following subsection we present an extension to the conditional independence approach, which may capture better the dependency structure of $z_{t,s,d}$ over $d$. 

\subsection{Extension of the conditional independence approach}\label{sec:extension}
We begin by noting that modelling $z_{t,s,d}$ with a Negative-Binomial distribution is equivalent to modelling $z_{t,s,d}$ as an over-dispersed Poisson quantity:
\begin{eqnarray}
 z_{t,s,d} \mid \mu_{t,s} &\sim& \mbox{Poisson}(\mu_{t,s,d})\\
 \mu_{t,s,d} &\sim & \mbox{Gamma}(\alpha_{t,s,d},\beta_{t,s,d}).
\end{eqnarray}
In this form we can consider the variance of $z_{t,s,d}$ as the sum of the variance of the Poisson component and the variance of the Gamma component. A Gamma component which contributes more to the total variance corresponds to a lower value for the Negative-Binomial shape parameter and vice-versa. In the GLM framework we assume that both the Poisson and Gamma quantities are conditionally-independent across the delay indices $d$. Noting that in Bayesian hierarchical modelling the Gamma component is often approximated by a Log-Normal component, where the mean at the log-level includes an identically distributed Normal random effect, one approach to modelling conditional covariance between multivariate counts is to model the Poisson mean $\bm{\mu}_{t,s}$ as a Multivariate-Log-Normal quantity \citep{poisson-log-normal}:
\begin{eqnarray}
z_{t,s,d} &\sim & \mbox{Poisson}(\mu_{t,s,d}) \\
\log(\bm{\mu}_{t,s}) & \sim & \mbox{Multivariate-Normal}(\bm{\nu}_{t,s},\bm{\Sigma}_{t,s})\\
\nu_{t,s,d} &=& f(t,s) + g(t,s,d)
\end{eqnarray}

In this framework, which we refer to as the ``GLM+ framework'', the partial counts $z_{t,s,d}$ are still independent given $\mu_{t,s,d}$. However, at least some of the total covariance can be described explicitly by the Multivariate-Normal covariance structure. The implication of this is that the model may be better able to capture the covariance structure of the $z_{t,s,d}$, and consequently the variance of the total counts $y_{t,s}$, compared to the GLM framework.

In the following section, we present a general framework based on the Multinomial mixture approach, which retains the desirable merits of jointly modelling $z_{t,d,s}$ as well as the necessary flexibility to capture variability in the spatio-temporal and delay structures.

\section{Generalized-Dirichlet-Multinomial Framework}\label{sec:framework}
Recall that $y_{t,s}$ denotes the true count occurring at temporal unit $t \in T$ and in spatial unit $s \in S$ and that $z_{t,s,d}$ denotes the observed count corresponding to $y_{t,s}$ with delay $d \in \{1,...,D\}$. We begin by defining a Negative-Binomial model for the true counts:
\begin{eqnarray}
y_{t,s} \mid \lambda_{t,s}, \theta &\sim & \mbox{Negative-Binomial}(\lambda_{t,s},\theta) \\
\log(\lambda_{t,s}) &=& f(t,s)
\end{eqnarray}
with $f(t,s)$ the same as in Section \ref{sec:background}. Modelling $y_{t,s}$ directly (as opposed to indirectly using the GLM), reduces the risk that Var($y_{t,s}$) will not be captured well. However in order to make predictions about $y_{t,s}$ which have not yet been fully observed, we also need a model for the delayed counts $\bm{z}_{t,s}$ (which should provide partial information on the unobserved $y_{t,s}$):
\begin{eqnarray}
\bm{z}_{t,s} \mid \bm{p}_{t,s},y_{t,s} &\sim & \mbox{Multinomial}(\bm{p}_{t,s},y_{t,s}).
\end{eqnarray}

Unlike the GLM approach, modelling the $\bm{z}_{t,s}$ in this way implies they are not assumed to be conditionally independent. In the simplest formulation of this framework, the $\bm{p}_{t,s}$ are not random but fixed, given any spatio-temporal structures or relevant covariates. However, this carries the risk of falsely confounding variability in the delay mechanism with variability in the true count $y_{t,s}$ when now-casting. We illustrate this by considering that the predictive distribution for unobserved totals $y_{t,s}$, conditional on partial counts $\bm{z}_{t,s}$, is given by:
\begin{equation}
p(y_{t,s}|\bm{z}_{t,s})\propto p(\bm{z}_{t,s}|y_{t,s})|p(y_{t,s})
\end{equation}
The issue is that $p(\bm{z}_{t,s}|y_{t,s})$ is Multinomial, which lacks flexibility in the variance as, conditional on $y_{t,s}$, both the mean and variance are defined wholly by $\bm{p}_{t,s}$. As such, if there is excess variability (over-dispersion) in $\bm{z}_{t,s}|y_{t,s}$, this is likely to be erroneously absorbed by $p(y_{t,s})$. For example, if $z_{t,s,1}/y_{t,s}$ is too high for the Multinomial to reasonably capture given $p_{t,s,1}$, predictions of $y_{t,s}$ may be too high when now-casting. Moreover, if both the mean and correlation structure $\bm{z}_{t,s}|y_{t,s}$ are exclusively defined by $\bm{p}_{t,s}$, then this limits flexibility in capturing unusual covariance structures.

Both of these issues can be addressed by modelling $\bm{p}_{t,s}$ as a $\mbox{Generalized-Dirichlet}(\bm{\alpha}_{t,s},\bm{\beta}_{t,s})$ distribution, the probability density function of which is: 
\begin{eqnarray}
p(p_1,p_2,...,p_k\mid \bm{\alpha},\bm{\beta}) = p_k^{\beta_{k-1}-1} \prod_{i=1}^{k-1} \left[ \frac{p_i^{\alpha_i-1}}{B(\alpha_i,\beta_i)}   \left(\sum_{j=i}^k p_j \right)^{\beta_{i-1}-(\alpha_i+\beta_i)} \right].  \label{GD}  
\end{eqnarray}

The resulting marginal model can be obtained by integrating out $\bm{p}_{t,s}$ to obtain a Generalized-Dirichlet-Multinomial or  GDM$(\bm{\alpha}_{t,s},\bm{\beta}_{t,s},y_{t,s})$ mixture distribution for $\bm{z}_{t,s}|y_{t,s}$, with probability mass function:
\begin{equation}
p(z_1,z_2,...,z_k\mid \bm{\alpha},\bm{\beta},y) = \frac{\Gamma(y+1)}{\Gamma(z_k+1)}\prod_{i=1}^{k-1} 
\left[ \frac{\Gamma(z_i+\alpha_i)\Gamma(\sum_{j=i+1}^k z_j + \beta_i)}{B(\alpha_i,\beta_i)\Gamma(z_i+1)\Gamma(\alpha_i+\beta_i+\sum_{j=i}^k z_j)} \right]. \label{GDM}
\end{equation}

To be useful as a tool for nowcasting and forecasting, the model needs to be able to provide inference for $y_{t,s}$ conditional on any corresponding $z_{t,s,d}$ which have been observed (as well as any preceding $y_{t,s}$ which have been observed by the time of prediction). In a Markov Chain Monte Carlo implementation framework (such as the one used here) this is possible by sampling the corresponding $z_{t,s,d}$ which have not yet been observed as well as the unobserved $y_{t,s}$. However, to do the former we need to be able to sample from the conditional distributions $z_{t,s,d}\mid z_{t,s,1},...,z_{t,s,d-1},y_{t,s}$. Fortunately, we can do this by defining and implementing the model in terms of the conditional structure of the GDM:
\begin{eqnarray}
z_i\mid \bm{z}_{-i}, \bm{\alpha}, \bm{\beta},y &\sim & \mbox{Beta-Binomial}(\alpha_i,\beta_i,n_i=y-\sum_{j< i} z_j) \label{dr:eq:beta-binomial}\\
p(z_i\mid \bm{z}_{-i}, \bm{\alpha}, \bm{\beta},y) &=& \binom{n_i}{z_i} \frac{B(z_i+\alpha_i ,n_i-z_i+\beta_i)}{B(\alpha_i , \beta_i)}.
\end{eqnarray}

To model structured variability in the delay mechanism, it makes sense to re-parametrise the Beta-Binomial in terms of its mean $\nu_{t,s,d}$ and dispersion parameter $\phi_{t,s,d}$, which relate to the parameters of the GDM by:
\begin{equation}
\alpha_{t,s,d} = \nu_{t,s,d} \phi_{t,s,d}; \qquad \beta_{t,s,d} = (1-\nu_{t,s,d}) \phi_{t,s,d} 
\end{equation}
The intuition behind this characterisation is that, having already observed some delayed counts $z_{t,s,1},...,z_{t,s,d-1}$ corresponding to the true count $y_{t,s}$, then $\nu_{t,s,d}$ represents the proportion of the remaining (so far unreported) counts we expect to be reported in the next delay step $d$. Variability about $\nu_{t,s,d}$ is controlled by the dispersion parameter $\phi_{t,s,d}$. Both the mean and dispersion parameters can be generally characterised as functions of space, time and delay:
\begin{eqnarray}
\log\left(\frac{\nu_{t,s,d}}{1-\nu_{t,s,d}}\right) &=& g(t,s,d) \\
\log(\phi_{t,s,d}) &=& h(t,s,d)\label{dr:eq:log_phi}.
\end{eqnarray}

In contrast to the GLM approach, predictive inference for the unobserved $y_{t,s}$ is based on both the delayed counts $\bm{z}_{t,s}$ and previous observed values $y_{t',s}$ for $t'<t$. In practice, using MCMC for model inference automatically generates predictive samples of the unobserved $y_{t,s}$ from $y_{t,s}|\bm{z}_{t,s},y_{t',s}$. Furthermore, the delay mechanism does not appear in the model for $y_{t,s}$, meaning that associated variability does not propagate into the predictive inference for unobserved $y_{t,s}$. In the subsequent section we will apply equivalent GDM, GLM and GLM+ models to dengue fever data, discussing their relative merits with respect to performance in model checking, now-casting and forecasting. 

\section{Case Study}\label{sec:comparison}
Dengue fever is a viral infection, transmitted by mosquitoes, which has flu symptoms that may evolve into a potentially fatal condition known as severe dengue \citep{who_dengue}. The disease causes a major burden for the population it affects, particularly in Brazil, which reports more dengue cases than any other country \citep{Silva2016AccuracyOD}. Effective response to dengue requires early detection \citep{who_dengue}, so it is important that healthcare providers are able to prepare themselves for a possible outbreak. Though the reporting of dengue cases to the Brazilian national surveillance system (SINAN) is mandatory \citep{Silva2016AccuracyOD}, it can take weeks or even months of delay for the number of reported cases occurring in a given week to approach a final count. For this reason, statistical delayed-reporting models are used to correct delays and predict outbreaks before the total count is available \citep{theo_dengue}.
\begin{figure}
\floatbox[{\capbeside\thisfloatsetup{capbesideposition={right,center},capbesidewidth=0.4 \linewidth}}]{figure}[1.2\FBwidth]
{\caption{Total number of reported dengue cases from 2011 onwards in Rio de Janeiro. Different colours represent which data are fully observed, partially observed or unobserved at week $t=114$ (March 2013). }\label{fig:dr:total_data}}
{\includegraphics[width=\linewidth]{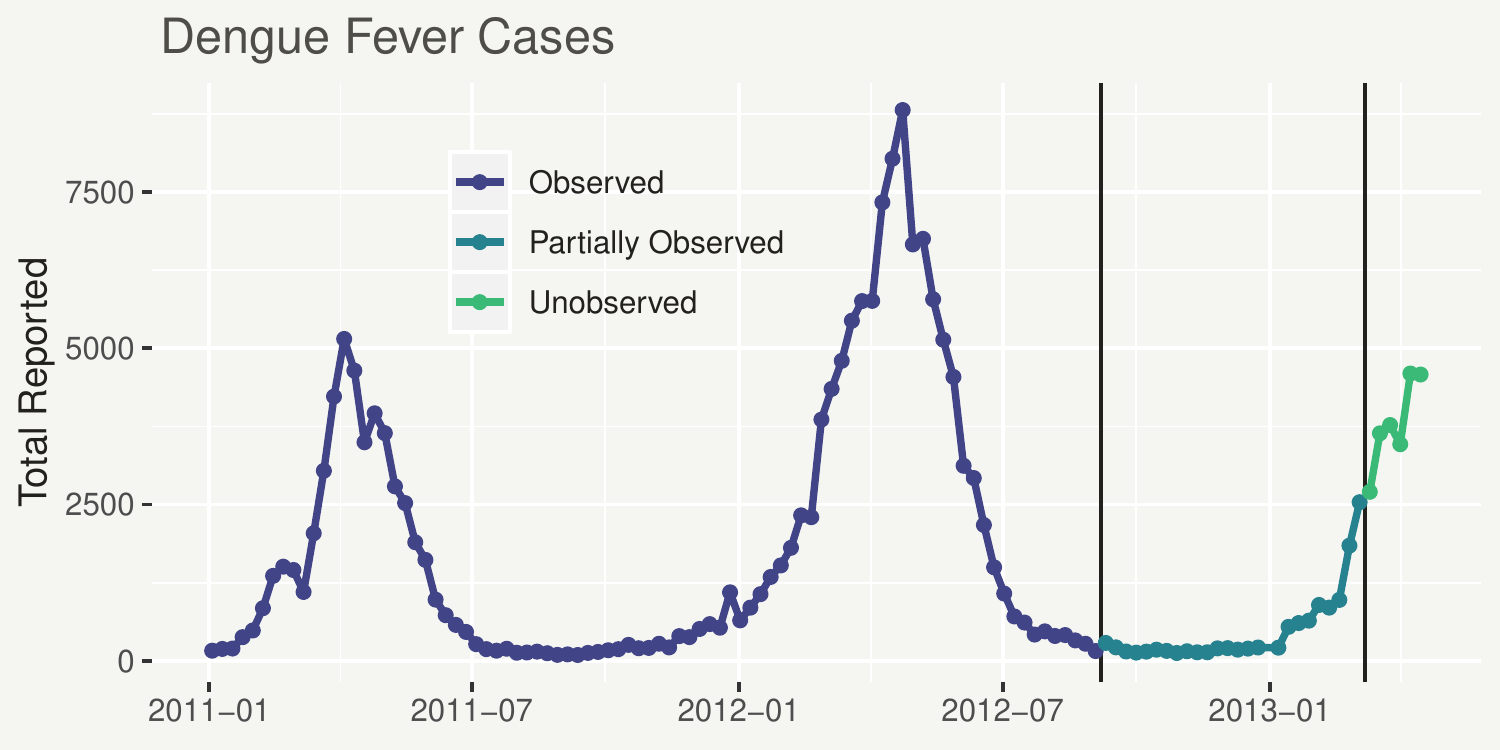}}
\end{figure}

Here we consider data on dengue fever cases in Rio de Janeiro, Brazil, occurring in weeks $t=1$ (week commencing the 3rd of January 2011) to $t=120$ (week commencing the 15th of April 2013). For illustration, we assume that present day is week $t=114$ (week commencing the 4th of March 2013). Furthermore, we consider the total observable count to be the number of cases observed after 6 months (26 weeks) worth of data (in addition to the number of cases reported in the week of occurrence). With present day being week $t=114$, this implies we have 88 weeks of fully observed total counts $y_t$. Total counts occurring in weeks $t=89$ to $t=114$ are only partially observed and must be predicted based on the partial observations (now-casting). Total counts $y_t$ after present day ($t=114$) have not yet occurred and so they are completely unobserved. This is the forecasting period.

The time series of counts is illustrated in Figure \ref{fig:dr:total_data}, with different colours corresponding to the three different periods. There is some evidence of seasonality in the data, with outbreaks starting around the beginning of the calendar year and ending approximately 6 months later. This reflects the fact that the incidence of dengue fever is thought to depend heavily on the time of year and climatological conditions \citep{dengue_seasonal}. We can also see some non-seasonal temporal structure, meanwhile, with the outbreak in 2012 being more severe than the one in 2011. 
\begin{figure}[h!]
\includegraphics[width=\linewidth]{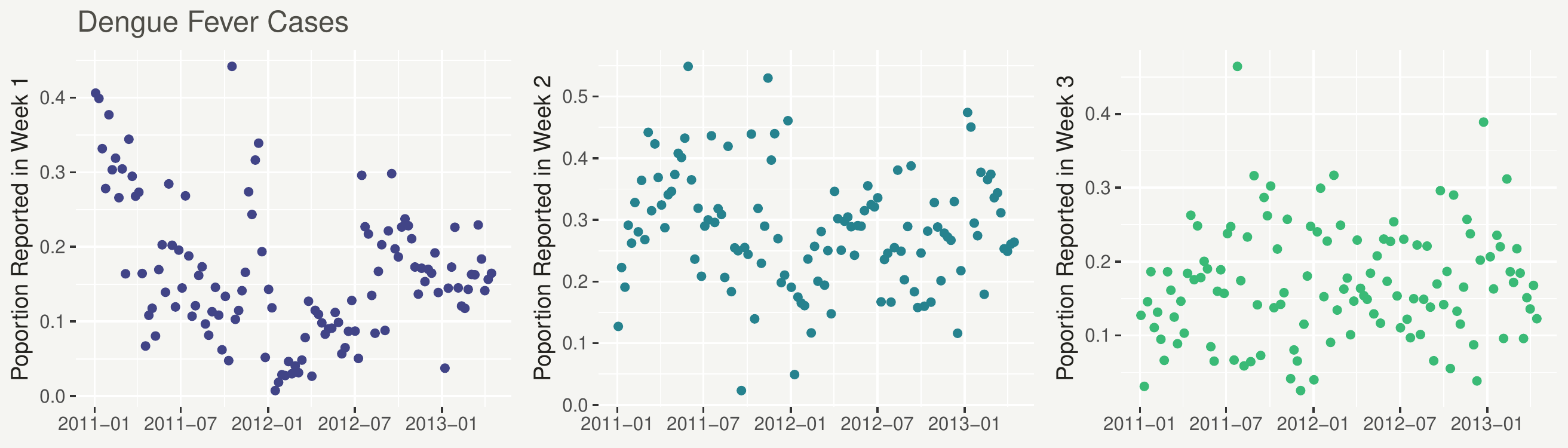}
\caption{Proportion of dengue cases reported in the first week (the week in which they occurred, left), the second week (centre) and the third week (right).}
\label{fig:dr:proportion_data}
\end{figure}

The left panel in Figure \ref{fig:dr:proportion_data} shows the proportion of dengue cases reported in the week they occurred (first week) plotted against time, while the middle and right panels show the proportion of cases reported a week after they occurred (second week) and the following (third) week, respectively. We can see strong evidence of temporal structure in the delay mechanism, with the average proportion reported in the first week steadily dropping throughout 2011, reaching its lowest point at the start of 2012 before beginning to rise again.

\subsection{Formulation of competing models}\label{sec:twomodels}
We now model this data using three comparable models (in terms of flexibility and interpretation), namely the GDM, GLM and GLM+. Modelling every partial count $z_{t,d}$ (in this case all 27 weeks) will result in the greatest predictive precision, though this comes at a high computational cost. Instead, if the total $y_t$ is almost entirely observed after $D$ delay steps, it may be more pragmatic to model only counts $z_{t,d}$ up to $d=D$ as well as the sum of the remaining counts $z_{t,D+1}=y_t - \sum_{d=1}^D z_{t,d}$. In the GDM approach this is achieved by only including the conditional models for the first $D$ partial counts, such that the remainder is modelled implicitly, while in the GLM and GLM+ approaches this can be achieved by modelling $z_{t,D+1}$ in the same way as the individual counts.
\begin{figure}
\floatbox[{\capbeside\thisfloatsetup{capbesideposition={right,center},capbesidewidth=0.4 \linewidth}}]{figure}[1.2\FBwidth]
{\caption{Quantiles of the proportions of fully observed ($t=1,...,88$) total dengue cases $y_t$ covered by $\sum_{d=1}^D z_{t,d}$ after each additional week of data.}\label{fig:dr:proportion_reported}}
{\includegraphics[width=\linewidth]{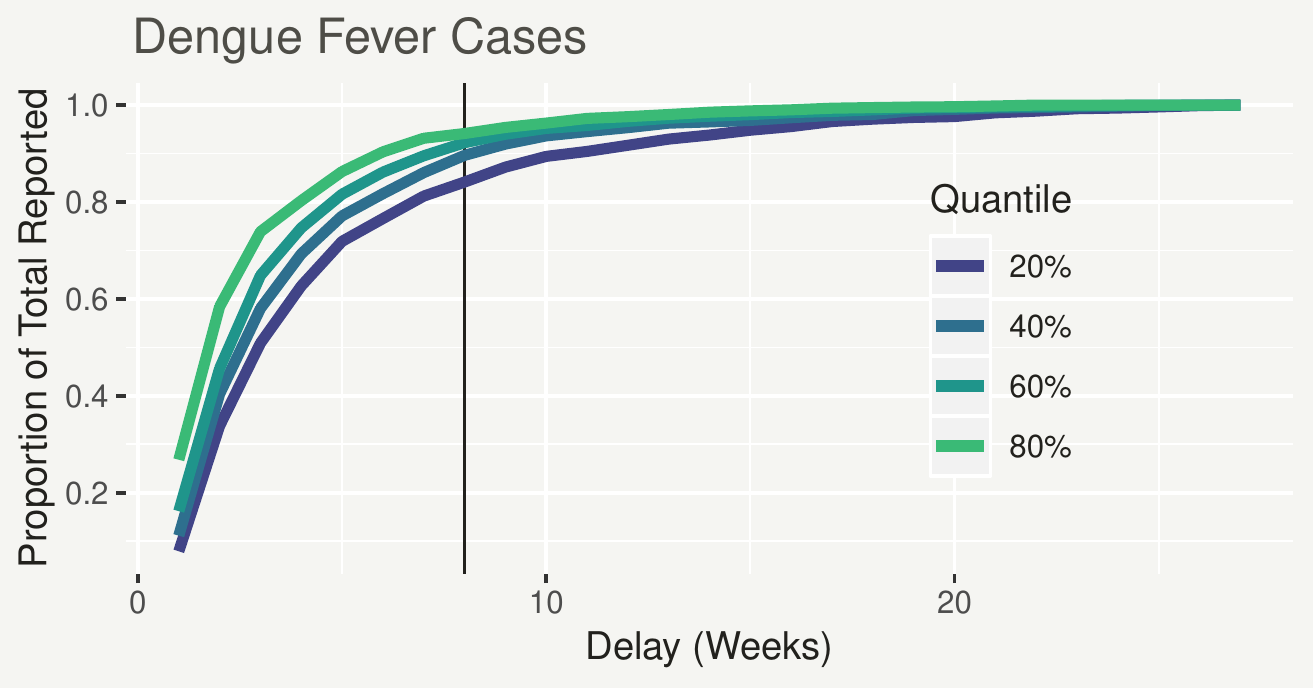}}
\end{figure}

One way to make this decision is to consider the proportions of each observed $y_t$ reported after each delay step. Figure \ref{fig:dr:proportion_reported} shows the 20\%, 40\%, 60\%, and 80\% quantiles of the proportions of the total dengue cases reported after each delay step. By looking at the 20\% quantiles of these proportions we can see that the vast majority (over 80\%) of total dengue cases are covered after  $D=8$ weeks worth of data 80\% of the time, with little to be gained unless many more weeks are considered. For this reason we choose to model only the first 8 weeks individually.

The model based on the GDM framework is defined by:
\begin{eqnarray}
y_t &\sim& \mbox{Negative-Binomial}(\lambda_t,\theta) \\
\log(\lambda_t) &=& \iota + \alpha_t + \eta_t \\
\bm{z}_t \mid y_t &\sim& \mbox{GDM}(\bm{\nu}_t,\bm{\phi},y_t) \\
\log\left(\frac{\nu_{t,d}}{1-\nu_{t,d}}\right) &=& \psi_d + \beta_{t,d}
\end{eqnarray}
Where $\bm{\nu}_t$ and $\bm{\phi}$ are the expectations and dispersions parameters of the Beta-Binomial conditional distributions, as described in \eqref{dr:eq:beta-binomial}-\eqref{dr:eq:log_phi}.

The model based on the GLM framework is defined by:
\begin{eqnarray}
z_{t,d} &\sim& \mbox{Negative-Binomial}(\mu_{t,d},\theta_d) \\
\log(\mu_{t,d}) &=& \iota +\alpha_t + \eta_t + \psi_d + \beta_{t,d} 
\end{eqnarray}

The model based on the GLM+ framework is defined by:
\begin{eqnarray}
z_{t,d} &\sim& \mbox{Negative-Binomial}(\mu_{t,d},\theta_d) \\
\log(\bm{\mu}_t) &\sim & \mbox{Multivariate-Normal}(\bm{\nu_t},\bm{\Sigma}) \\
\nu_{t,d} &=&\iota + \alpha_t + \eta_t + \psi_d + \beta_{t,d} 
\end{eqnarray}

In all models $\eta_t$ is a penalized cyclic cubic spline \citep{GAM} defined over weeks $1,...,52$, which represents the effect of the time of year on the total number of reported dengue cases, and $\alpha_t$ is a penalized cubic spline defined over the whole temporal range. The latter is designed to capture non-seasonal temporal structure in the rate of total reported dengue cases and is constrained to be linear beyond the final knot so that it can be used for forecasting. The effects $\beta_{t,d}$ are defined by a different penalized cubic spline (each with its own smoothness penalty) for each delay index $d$, intended to capture temporal changes in the delay mechanism over time. As discussed in \cite{JAGAM}, the coefficients of each spline are assigned a Multivariate-Normal prior distribution and are penalized to prevent excessive wiggliness through an unknown penalty parameter $\tau$ (the scaling factor of the Multivariate-Normal prior precision matrix). The re-parametrisation $\sigma=1/\sqrt{\tau}$ is potentially more interpretable for the purpose of specifying a prior distribution, where smaller values of $\sigma$ correspond to a stricter penalty on how flexible the smooth function is. The splines are centred to have zero-mean, and as such the models include the fixed effects $\iota$ and $\psi_d$ as intercepts. 

The Negative-Binomial dispersion parameters ($\theta_d$ and $\theta$) were assigned relatively non-informative Exponential($0.01$) prior distributions. The GDM dispersion parameters $\phi_d$ were assigned Log-Normal($2,2$) prior distributions, such that most of the prior density is over values of $\phi_d$ which result in a modest contribution from the Generalized-Dirichlet component to the overall variance of the GDM, without ruling out higher values which correspond to a Multinomial situation. Relatively non-informative Normal$(0,10^2)$ prior distributions were specified for the global intercept parameter $\iota$ and also for the delay-specific intercept parameters $\psi_d$. In the GDM model, the intercept parameters $\psi_d$ represent the means of relative proportions at the logistic level. For these parameters we specified Normal prior distributions with the means chosen so that the prior mode implies approximately equal amount of cases being reported in each week of delay, with the variance chosen so that they are relatively non-informative. We specified Half-Normal$(0,1)$ prior distributions for the penalty parameters for splines $\alpha_t$ and $\eta_t$. This imposes a relatively strong smoothness penalty on the effects $\alpha_t$ and $\eta_t$, which are supposed to capture medium-to-long term trends in the incidence of dengue cases. We relaxed this penalty slightly for the effects $\beta_{t,d}$ by specifying weaker Half-Normal$(0,\sqrt2)$ priors. Finally, for the Multivariate-Normal covariance of $\log(\bm{\mu}_t)$ in the GLM+ model, we specified a fairly weak Inverse-Wishart prior with an identity scale matrix (dimension $D+1$) and $D+2$ degrees of freedom.

All code was written and executed using R (\cite{R}) and all three models were implemented using NIMBLE \citep{nimble}, a facility for highly flexible implementation of MCMC. The model matrices for the splines were set up using the package \texttt{jagam} \citep{JAGAM}. Four MCMC chains were run from different initial values and with different random number generator seeds, until convergence criteria were met. We discuss how we assessed convergence of the chains to the posterior in the Appendix \ref{dr:app:convergence}. 

\subsection{Results}
\begin{figure}[h!]
\includegraphics[width=\linewidth]{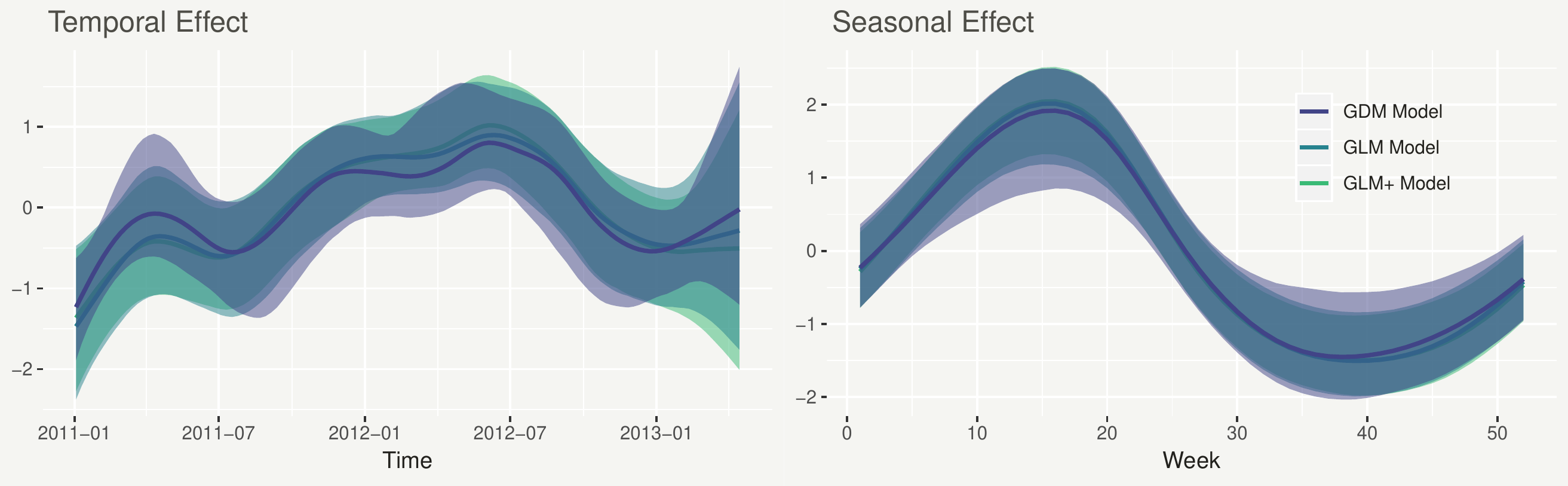}
\caption{Posterior mean temporal ($\alpha_t$) and seasonal ($\eta_t$) spline effects on the incidence rate of dengue cases, from the GDM, GLM and GLM+ models, with associated 95\% credible intervals.}
\label{fig:total_effects}
\end{figure}

To compare the models we will begin by exploring which aspects of the results are similar. Figure \ref{fig:total_effects} shows the posterior mean predicted temporal effect ($\alpha_t$) as well as the seasonal effect ($\eta_t$) from the GDM, GLM and GLM+ models, with associated 95\% credible intervals, on the incidence rate dengue cases (at the log-scale). Both effects are very similar in shape between the three models: in the left panel we can see that all models suggest a persistent increase in dengue incidence in 2012, which makes sense given the more severe outbreak shown in Figure \ref{fig:dr:total_data}, while the right panel shows a strong seasonal effect in all models, with a much higher incidence rate in the first half of the year than the second. Interestingly the seasonal effect is less certain, though still strong, for the GDM model compared to the GLM and GLM+ models. Given that there are only approximately two years of fully observed data, the uncertainty in the GDM model's seasonal effect seems more reasonable. 

\begin{figure}
\floatbox[{\capbeside\thisfloatsetup{capbesideposition={right,center},capbesidewidth=0.5 \linewidth}}]{figure}[1\FBwidth]
{\caption{Posterior mean delay spline effect $\beta_{t,1}$ corresponding to counts reported in the first week $z_{t,1}$, from the GDM, GLM and GLM+ models, with associated 95\% credible intervals.}\label{fig:dr:delay_1}}
{\includegraphics[width=\linewidth]{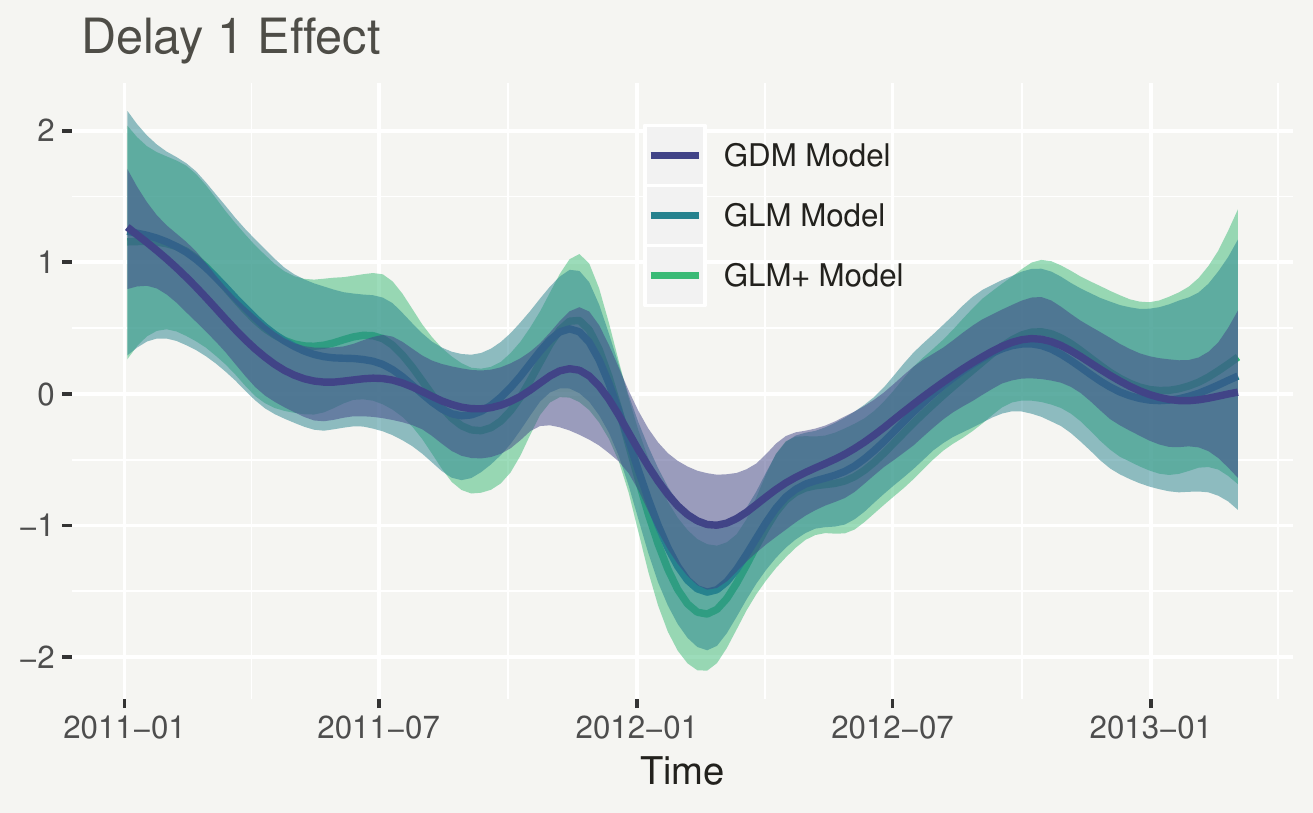}}
\end{figure}

Similarly, Figure \ref{fig:dr:delay_1} shows that, although not perfectly comparable because the models use different link functions (logistic for GDM and log for GLM and GLM+), the temporal effects on the number of cases reported in the first week are very similar between the three models. For example, all three models show a distinct drop in proportion of cases reported in the first week during the 2012 outbreak.

We now move on to ways in which the models differ. Recall from Section \ref{sec:background}, that in the GLM framework, capturing the distribution of the true counts $y_{t,s}$ well relies on a potentially restrictive assumption that the delayed counts $z_{t,s,d}$ are conditionally independent. In contrast, by modelling the total counts $y_{t,s}$ explicitly, the GDM framework has more flexibility to capture their distribution well. Similarly, the addition of a covariance model in the GLM+ framework means that it may be able to capture the covariance of the partial counts $z_{t,d}$, and consequently the variance of the total counts, better than the GLM framework.

We use in-sample posterior predictive checking \citep{Gelman2013} to the fit of the models to the data. This is done by simulating replicates of the observed partial counts $\tilde{z}_{t,d} \mid z_{t,d}$ and the fully observed (weeks 1-88) total dengue counts $\tilde{y}_t \mid y_t$ from the respective predictive distributions. We can then see if particular statistics of the observed data are captured well, by comparing them to the distribution obtained by computing the corresponding statistics of the replicates. 

\begin{figure}
\floatbox[{\capbeside\thisfloatsetup{capbesideposition={right,center},capbesidewidth=0.33 \linewidth}}]{figure}[1.33\FBwidth]
{\caption{Density plots of the mean bias (left column) and the logarith of the mean squared error (right column) of the covariance of the partial counts $z_{t,d}$ and the proportion reported in each week $z_{t,d}/y_t$.}\label{fig:dr:cov_checks}}
{\includegraphics[width=\linewidth]{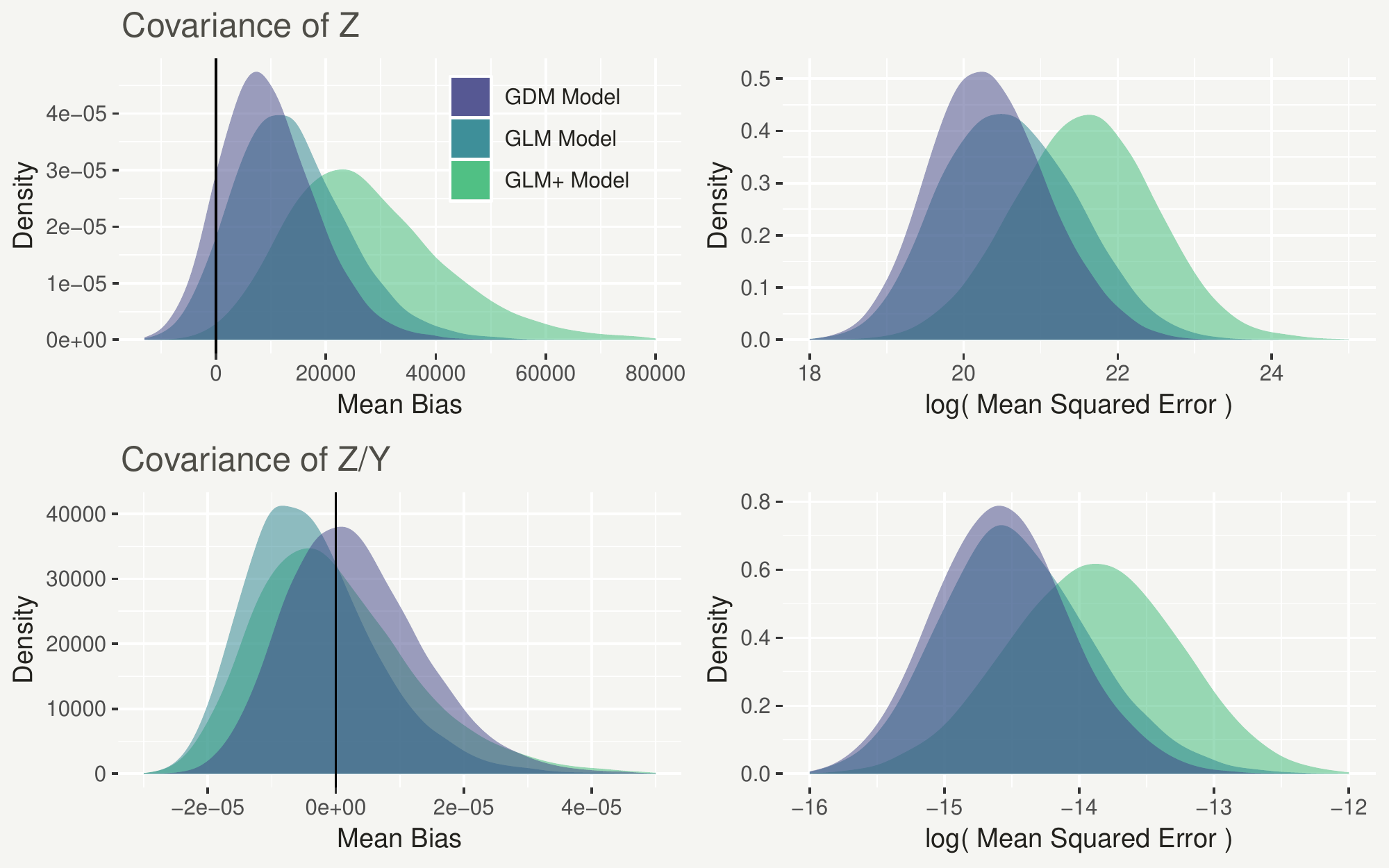}}
\end{figure}

We begin by looking at the covariance of the partial counts $z_{t,d}$ and the covariance of the proportion reported in each week $z_{t,d}/y_t$. For each sets of replicates, we compute the sample covariance of these two quantities, resulting in a distribution of samples for each individual covariance $\mbox{Cov}[\bm{\tilde{z}}_{i},\bm{\tilde{z}}_j]$ and $\mbox{Cov}[\bm{\tilde{z}}_{i}/\bm{\tilde{y}},\bm{\tilde{z}}_j/\bm{\tilde{y}}]$. The left column of Figure \ref{fig:dr:cov_checks} shows the mean difference between the replicate covariances and the observed covariances, while the right column shows the mean squared difference between the replicate covariances and the observed covariances. For both the covariance of the partial counts and the covariance of the proportion reported in each week, we can see that the GDM model is the least biased (potentially even unbiased for the proportion reported in each week) and the most precise (lowest mean squared error).

\begin{figure}[h!]
\includegraphics[width=\linewidth]{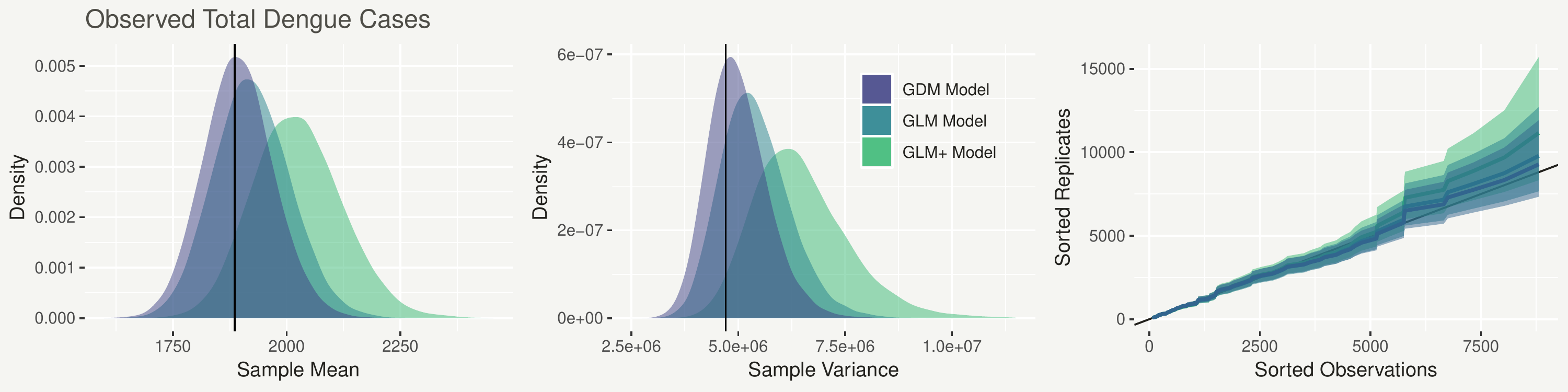}
\caption{The left and central panels show density plots of the sample mean and sample variance of the posterior replicates of the fully observed (weeks 1-104) total dengue cases ($y_{t}$) from the GDM and GLM models. The vertical lines represent the corresponding statistics from the observed data. The right panel shows the mean of replicates of the total dengue cases $y_{t}$, from the GDM and GLM models, with associated 95\% posterior predictive intervals.}
\label{fig:dr:checks}
\end{figure}

Similarly, the left and central panels of Figure \ref{fig:dr:checks} show density estimates of the distribution of the sample mean and the sample variance, respectively, of the replicate total counts $\tilde{y}_t.$ We can see that in both cases the observed statistic, represented by a vertical line, is captured best by the GDM model, with the GLM faring better than the GLM+ model. This is a surprising result, given that the GLM+ has more flexibility than the GLM to capture the covariance structure of the partial counts $z_{t,d}$. The right panel of Figure \ref{fig:dr:checks} shows posterior means of the sorted replicates, with 95\% prediction interval. In this plot we can clearly see that, while the distribution of the total counts is captured best by the GDM and adequately well by the GLM, the GLM+ has an excessively heavy upper tail, compared to the data. This difference is likely because in the Poisson-Log-Normal mixture the logarithm of the Poisson mean is symmetric, compared to negatively skewed in the Poisson-Gamma mixture.

\begin{figure}[h!]
\includegraphics[width=\linewidth]{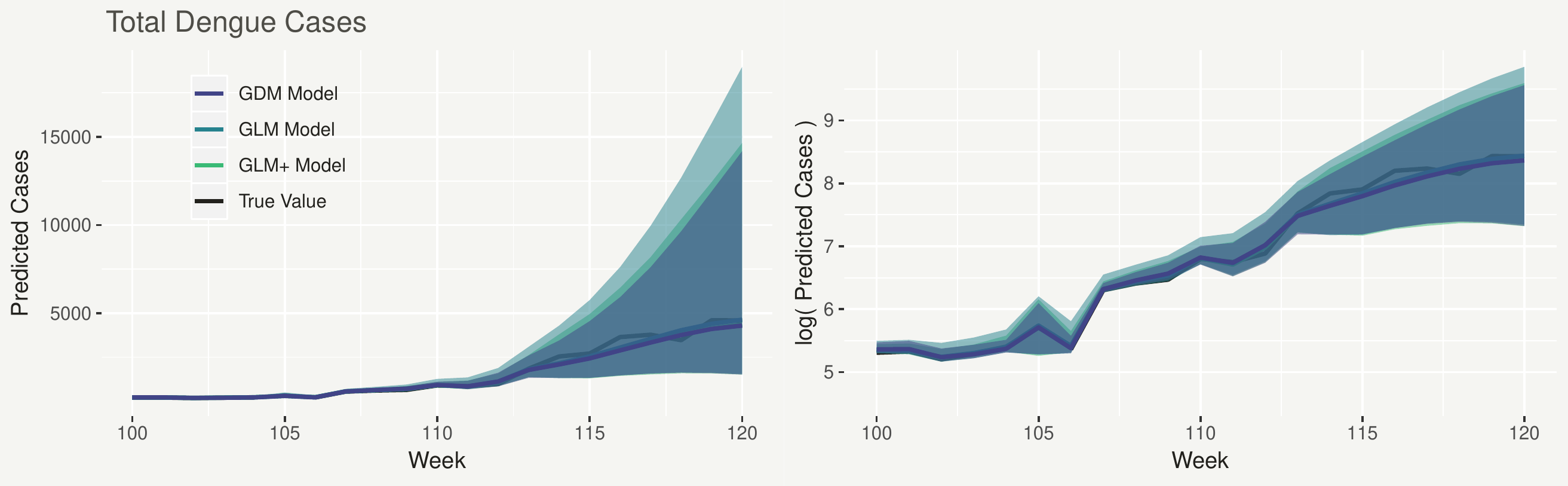}
\caption{Posterior median predictions of the unobserved total dengue cases $y_{t}$, from the GDM, GLM and GLM+ models, with associated 95\% posterior predictive intervals. Predictions beyond week $t=114$ are forecasting without any observed partial counts $z_{t,d}$.}
\label{fig:dr:forecasts}
\end{figure}

Recall that two important uses of delayed-reporting models are the prediction of total counts $y_{t,s}$ which have occurred but haven't yet been fully observed (nowcasting) and the prediction of total counts which have not yet occurred (forecasting). In this case study we imagine we are in week 114 and we would like to predict the number of dengue cases in recent weeks (e.g. $y_{114}$) as well as to predict dengue cases over the next 6 weeks. Figure \ref{fig:dr:forecasts} shows the posterior median predicted number of dengue cases $y_t$ from the three models, with associated 95\% posterior predictive intervals. We can see that, whilst the median predictions from all three models are virtually identical, the model with the least predictive uncertainty, in both the now-casting range and forecasting range, is the GDM, making the GDM forecast potentially most useful to decision-makers. Notably, the GLM+ is far closer to the GDM in terms of certainty than the GLM, suggesting our extension may have improved now-casting and forecasting precision. However, we would consider the GDM's quantification of uncertainty more trustworthy, given its favourable results in the in-sample predictive checking.

\subsection{Comparison with other approaches}
By this point we have demonstrated several ways, for this data, in which the GDM framework improves over the GLM framework our own extension of it, the GLM+ framework. It remains to show that the increased flexibility of the GDM over other approaches discussed in Secton \ref{sec:background} leads to tangible improvements in this example. Recall that one method presented by \cite{hohle2014bayesian} and others, is to treat the parameters of the Generalized-Dirichlet component as stationary in time. As we saw in Figure \ref{fig:dr:proportion_data}, there is substantial variation over time in the proportion of dengue cases reported in the first week. This structure would not be captured by assuming time-stationarity in the Generalized-Dirichlet model, inevitably leading to poorer nowcasting and forecasting performance.

An alternative suggestion was to model the proportion of cases reported at each delay level in each week using a conventional Multinomial logistic regression, removing the additional variability provided by the Generalized-Dirichlet component. 

\begin{figure}[h!]
\includegraphics[width=\linewidth]{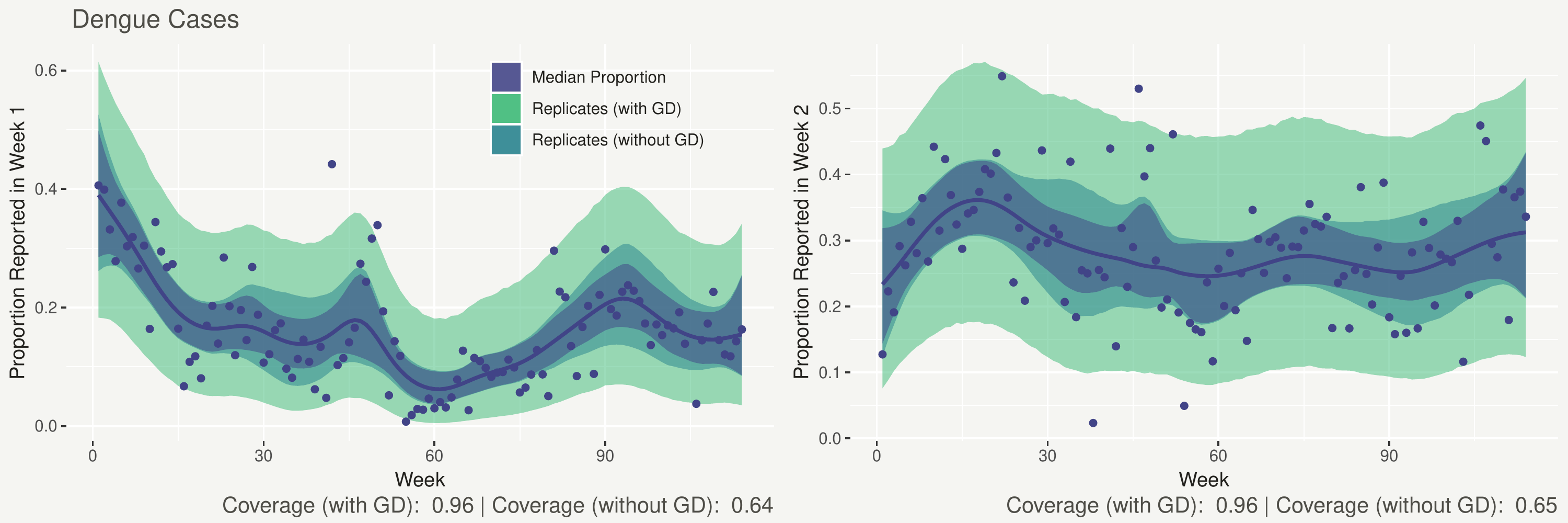}
\caption{Posterior median proportion, from the GDM model, of dengue cases reported in the first (left) and second (right) weeks after incidence, with associated 95\% credible intervals. Also shown are 95\% posterior predictive intervals of the proportion reported in the first and second weeks from the GDM model with and without the additional variance from the Generalized-Dirichlet layer.}
\label{fig:dr:weeks}
\end{figure}

One way to assess the contribution of the GD variance is to simulate posterior replicates of the proportion reported in each week of delay ($z_{t,d}/y_t$) both from the GDM using the posterior samples for the dispersion parameters $\phi_d$ and again from the same model but in the limiting case when $\phi_d \rightarrow \infty$, such that the joint conditional distribution of $z_{t,d}$ is Multinomial. Figure \ref{fig:dr:weeks} shows 95\% posterior predictive distributions for the proportion of dengue cases reported after 1 (left) and 2 (right) weeks of delay for both the model with GD variance and without. We can see that without the GD variance an excessively high number of points are not captured by the prediction intervals. Also shown are the 95\% prediction interval coverages: the proportion of observations which lie within their corresponding 95\% prediction intervals. The coverages with the GD variance are just over 95\%, indicating a good fit to this data, while less than two-thirds of points are covered without the GD variance.

\section{Under-reporting}\label{sec:under-reporting}
An added challenge that occurs in data that are subject to reporting delay is that, in some situations, the final observed total count $y_{t,s}$ may still be a (substantial) under-estimate of the true count. In disease surveillance, this may translate to many cases never being reported, leading to a biased understanding (underestimation) of the actual magnitude of outbreaks. For instance, although reporting of dengue cases to the national surveillance system (SINAN) is mandatory, research suggests that the reported total may be substantially lower than the true number of dengue cases, owing to issues such as patients not seeking healthcare \citep{Silva2016AccuracyOD}. 

To address this, the GDM framework can be adapted to allow for under-reporting. In particular, it can be integrated into the hierarchical framework for under-reporting presented in \cite{CUR}. Suppose that in addition to the partial counts $z_{t,s,d}$ and the total counts $y_{t,s}$ there exist unobserved true counts $x_{t,s}$ such that $y_{t,s}\leq x_{t,s}$. Then the complete model for delayed reporting and under-reporting is given by:
\begin{eqnarray}
x_{t,s} \mid \lambda_{t,s}, \theta &\sim & \mbox{Negative-Binomial}(\lambda_{t,s},\theta) \\
y_{t,s} \mid x_{t,s}, \pi_{t,s} & \sim & \mbox{Binomial}(\pi_{t,s},x_{t,s}) \\
\log\left(\frac{\pi_{t,s}}{1-\pi_{t,s}}\right) &=& i(t,s) \label{pi}\\
\bm{z}_{t,s}|y_{t,s} &\sim& GDM(\bm{\nu},\bm{\phi},y_{t,s}) 
\end{eqnarray}
such that $\lambda_{t,s}$ now represents the incidence rate of the true count $x_{t,s}$ (as opposed to the total observed count $y_{t,s}$) and $\pi_{t,s}$ represents the reporting rate. As illustrated in \cite{CUR}, both covariates and random effects can be used to model this reporting rate at the logistic level, represented by the generic function $i(t,s)$ in \eqref{pi}. 

Without any observations for the true count $x_{t,s}$, the model is not identifiable between a high reporting rate $\pi_{t,s}$ and a low incidence rate $\lambda_{t,s}$ or vice versa, but this can be resolved through the use at least one informative prior (such as for the overall reporting rate across the whole time series, as discussed in \cite{CUR}). 

Using this approach means that policy and intervention can be based on predictions for the true number of cases, taking into account both the delayed reporting and under-reporting mechanisms, to reduce the risk of an undersized response. Note further, that allowing for under-reporting in the total count would be much less straightforward using the GLM and GLM+ approaches, mainly because the totals $y_{t,s}$ are not modelled explicitly.

\section{Discussion}\label{sec:discussion}
In this article we have introduced the problem of delayed-reporting and its implications for society. We explained that it is a problem based around prediction, providing a motivation for a statistical approach to the problem. We explored several existing approaches, focusing on (a) approaches based on a Multinomial mixture distribution with either a time stationary Generalized-Dirichlet distribution or a logistic regression and (b) the conditional independence (GLM) approach. Both approaches are very flexible, in terms of incorporating complex spatio-temporal structures. However, we argue that they both have limitations: The approaches based on a Multinomial mixture are not sufficiently flexible to simultaneously capture delay mechanisms which are both heterogeneous in time and over-dispersed, with respect to the Multinomial variance. The GLM approach, on the other hand, does not explicitly model the total counts. This means it relies on capturing the covariance structure of the partial counts well in order to capture the distribution of the total counts well. This is hindered by the assumption that the partial counts are independent, conditional on any covariate or random effects. To potentially address this, we proposed an extension to this approach (which we refer to as the GLM+) which includes an explicit covariance model for the partial counts, with the aim of better capturing the distribution of the total counts.

We have proposed a general framework based on a Generalized-Dirichlet-Multinomial mixture, where the true total counts are explicitly modelled (unlike the GLM) and where the Multinomial probabilities are a Generalized-Dirichlet whose parameters may vary in space and time.  We presented a case study of data on reported dengue fever cases in Rio de Janeiro. In-sample predictive model checking was used to assess the models with respect to how well the distribution of the total number of cases was captured. Out-of-sample predictive checking was also used to assess performance when nowcasting and forecasting. We found that in every test the GDM has the strongest performance, even compared to the GLM+ model which, despite potentially having the most general covariance structure of the three models, was hindered by having an excessively heavy upper tail. 

In addition to considering the performance of each model for the particular data set, it is also important to consider other reasons why one might be preferable over the others. The GLM model, for instance, is by far the easiest to implement, especially in a non-Bayesian setting such as the Generalized Additive Model framework or in an approximate Bayesian setting such as INLA. The GDM, however, lends itself more to a full Bayesian treatment, where Markov Chain Monte Carlo (MCMC) is used, compared to the other frameworks. This is because the effects associated with the true total count and the effects associated with the delay mechanism are separated into different parts of the model and are related to different parts of the data (the total counts and the partial counts, respectively). In the GLM and GLM+ frameworks, meanwhile, all of the effects are in the same model and they end up competing with each other. For this reason, it is possible to obtain a higher effective number of posterior samples per second with the GDM model. 

In our view, the GDM framework is the most interpretable of all of the frameworks discussed here. This is because the delay mechanism, and any associated variability, is completely separated from the process which generates total counts. This makes in turn it easier to adapt the model for a given data problem. For example, we can see in Figure \ref{fig:dr:weeks} that the variability in the proportion of cases reported in the first week decreases in the middle of the time series. To capture this, it is a fairly trivial modification to model the logarithm of the dispersion parameters $\phi_{t,s,d}$, as defined in \eqref{dr:eq:log_phi}, using a penalized spline in time. Knowing that variability in the delay mechanism at a certain time is likely to be lower or higher than usual could further improve now-casting precision. Whilst it would be possible to model the Negative-Binomial dispersion parameters $\theta_d$ as time-varying in the GLM and GLM+ frameworks, there is no equivalent way of separating temporal structure in the variance of the total counts, from structure in the variance of the delay mechanism, as is possible in the GDM framework. On the same theme of adaptability, the GDM framework can also be easily integrated into a hierarchical framework for correcting under-reporting, which may be essential in scenarios where the final observed total count is still a substantial under-representation of the true count. In such situations, allowing for both the delay mechanism and the under-reporting mechanism simultaneously may be crucial for well-informed decision making.

\begin{appendices}
\section{Convergence of MCMC Chains}\label{dr:app:convergence}
For each model, convergence of the four chains was assessed by visual inspection of trace plots and by computing the Multivariate Potential Scale Reduction Factor (MPSRF) \citep{convergence} of a selection of model parameters. This compares the variance between the chains to the variance within the chains. If the two variances are similar then this typically results in an MPSRF of less than 1.05. Starting from different initial values and obtaining an MPSRF of less than 1.05 gives the best indication that the chains have converged to the posterior distribution.
\begin{itemize}
\item For the GDM model, we computed the MPSRF of every 10th $\lambda_t$ ($\lambda_{10},\lambda_{20},...$), $\theta$, every 10th $\beta_{t,d}$ and the $\phi_d$. The model was run for a total of 400k iterations, discarding the first 200k as burn-in and thinning by 20 to save memory. The MPSRF was computed to be 1.05 indicating that the model had converged. 
\item For the GLM model, we computed the MPSRF of every 10th $\mu_{t,d}$ and the $\theta_d$. The model was run for a total of 800k iterations, discarding the first 400k as burn-in and thinning by 40 to save memory. The MPSRF was computed to be 1.04.
\item For the GLM+ model, we computed the MPSRF of every 10th $\mu_{t,d}$. The model was run for a total of 800k iterations, discarding the first 400k as burn-in and thinning by 40 to save memory. The MPSRF was computed to be 1.02
\end{itemize}
\end{appendices}

\bibliographystyle{chicago}

\bibliography{library}
\end{document}